\documentclass[pre,twocolumn,showpacs,showkeys]{revtex4}
\usepackage{graphicx}% Include figure files
\usepackage{color}
\begin{document}

\title{Surface and bulk properties of ballistic deposition models with bond breaking}
\author{Juvenil S. Oliveira Filho, Tiago J. Oliveira and Jos\'e Arnaldo Redinz}
\affiliation{Departamento de F\'{i}sica, Universidade Federal de Vi\c{c}osa\\
36570-000, Vi\c{c}osa, MG - Brazil}

\date{\today}

\begin{abstract}
We introduce a new class of growth models, with a surface restructuring mechanism in which impinging particles may dislodge suspended particles, previously aggregated on the same column in the deposit. The flux of these particles is controlled through a probability $p$. These systems present a crossover, for small values of $p$, from random to correlated (KPZ) growth of surface roughness, which is studied through scaling arguments and Monte Carlo simulations on one- and two-dimensional substrates. We show that the crossover characteristic time  $t_{\times}$ scales with $p$ according to $t_{\times}\sim p^{-y}$ with $y=(n+1)$ and that the interface width at saturation $W_{sat}$ scales as $W_{sat}\sim p^{-\delta}$ with $\delta = (n+1)/2$, where $n$ is either the maximal number of broken bonds  or of dislodged suspended particles. This result shows that the sets of exponents $y=1$ and $\delta=1/2$ or $y=2$ and $\delta=1$ found 
in all previous works focusing on systems with this same type of crossover are not universal. Using scaling arguments, we show that the bulk porosity $P$ of the deposits scales as $P\sim p^{y-\delta}$ for small values of $p$. This general scaling relation is confirmed by our numerical simulations and explains previous results present in literature.
\end{abstract}

%\pacs{81.15.Aa 05.40.-a 68.55.-a 68.35.Ct}
%\keywords{...}
\maketitle

%\newpage

\section{Introduction}

The morphology and evolution of the surface and porous structure in thin-film growth has been a subject of extensive theoretical and experimental interest \cite{Barab,Meak}. From a theoretical side, a number of surface growth models have been proposed, one of the first being the ballistic deposition (BD) model \cite{Vold}. The BD model was initially formulated to explain sedimentary rock formation, but has also been extensively studied as a model of low-temperature thin-film growth and surface roughening \cite{Meak}. In the last few decades, the surface properties of the BD model were widely studied (see refs. \cite{Barab,Meak,DBS} and references therein) and it is believed to be in the Kardar-Parisi-Zhang (KPZ) \cite{KPZ} universality class. More recently, some efforts were also devoted to understanding its bulk structure \cite{Kons,Robledo}. Furthermore, some variations of the BD model were also considered. Some examples are ballistic deposition with oblique incidence of particles
 \cite{TMLu,YuAmar1}, slippery BD \cite{Sink,YuAmar2,Robledo}, and competitive growth models in which particles are deposited following probabilistically either BD or other deposition rules \cite{Pelle,Mura,Kola,Horo0,Silva,Chame,Brau,FDA2,Horo1}.

In all these models and in all limited mobility models, after deposition, the particles are permanently aggregated in the deposit and do not change their positions. However, in many common thin film deposition techniques such as sputter and pulsed laser deposition \cite{PVD} high energy particles could impinge the film and restructuring mechanisms at surfaces could be operative. Also in epitaxial growth, at low temperatures, some impinging atoms do not immediately dissipate the energy released upon formation of the atom-surface bond and, thus, these ``hot'' atoms may start some transient movements at the surface. Some examples are \textit{downward funneling}, \textit{transient mobility}, \textit{knockout} and \textit{cascading knockdown} of adatoms, which are processes responsible for smoothing the surface in low temperatures \cite{Evans0,Evans1,Voter}. In the last one, the ``hot'' atoms dislodge previously deposited atoms that are only partially supported (suspended atoms), and they can move together to the ``bottom valley''. Such a mechanism induces the formation of a more compact film, eliminating overhangs at the surface by means of restructuring of deposited particles.

In order to study the effects of restructuring mechanisms on surfaces, we introduce here a different class of BD-like models considering deposition of particles which reach the surface with different energies. The low energy particles stick at the first contact with the surface, as in the plain BD model, and are deposited with probability $p$. On the other hand, high energy particles are deposited in a ``random deposition-like'' (RD-like) way, with probability $1-p$. When this kind of particle impinges over (previously deposited) suspended ones, their bonds can break and these particles can fall together until reaching the next top of the column. We model this restructuring mechanism in two different ways: considering that up to $n$ bonds can break (BBBDB models) or that up to $n$ suspended particles at the same column can change their positions, independently of the number of lateral bonds (BBBDS models). In both cases, $n$ is associated with the energy of the incident particle, as the larger this energy is the larger is the effect produced by it at the deposit and, thus, the larger is the number of broken bonds or particles dislodged. When bonds do not break, the ``RD-like'' particle is deposited on the top of the column (simple RD rule).

Since for $p=0$ and $p=1$ the BBBD models behave as plain RD and BD models, respectively, they present a competition from random to correlated (KPZ) growth. At short times the surfaces grow as in uncorrelated systems, its roughness (or width) $W$ scales in time as $W\sim t^{1/2}$. In a crossover time $t_{\times}$, which scales with system size and $p$ as $t_{\times} \sim p^{-y} L^{z_{BD}}$, correlations develop and the roughness starts to scale as $W \sim p^{-\gamma} t^{\beta_{BD}}$. Finally, in a saturation time $t_S$, the roughness becomes constant in a value $W_{sat}$, which scales with the substrate size and $p$ as $W_{sat}\sim p^{-\delta} L^{\alpha_{BD}}$. The scaling exponents $\alpha_{BD}$, $\beta_{BD}$, $z_{BD}$, $y$, $\delta$ and $\gamma$ are not all independent, following the scaling laws: $z_{BD} = \alpha_{BD} / \beta_{BD}$ \cite{Barab} and $\gamma = \delta - y \beta_{BD}$ \cite{Horo0}. All these behaviors could be resumed in the compact form \cite{Horo0,Horo1}:
\begin{equation}
W\sim\frac{L^{\alpha_{BD}}}{p^{\delta}}F\left(\frac{t/p^{-y}}{L^{z_{BD}}}\right)\;\;\;\;\; (p\rightarrow 0),
\label{EqScaling}
\end{equation}
where $F(u)$ is a scaling function which behaves as $F(u)\sim u^{\beta_{BD}}$ for $u<<1$ and $F(u)\sim$ const. for $u>>1$. $\alpha_{BD}$ is the roughness, $\beta_{BD}$ is the growth and $z_{BD}$ is the dynamical exponents of the BD model. In general, $\delta$ and $y$ are exponents dependent on the correlated model and independent of the dimensionality of the system. Results for several competitive models \cite{Kola,FDA2,Kola2,Horo1} suggest that the exponents $\delta$ and $y$ are not directly linked to the universality class of the correlated process and can be separated into two groups: the first one including solid-on-solid (SOS) correlated models, for which $\delta=1$ and $y=2$; and the second one including correlated models with lateral aggregation (ballistic-like models), for which $\delta=1/2$ and $y=1$. These pairs of exponents are claimed to be universal, i.e., every model with a crossover from random to correlated growth would present one of those pairs of exponents \cite{Horo1}. In the BBBD models, we will show that the restructuring mechanism does not change the asymptotic class of the model, which is KPZ. However, it changes the exponents $\delta$ and $y$ that assume the general values $\delta=(n+1)/2$ and $y=n+1$, where $n$ is the maximal number of either broken bonds or dislodged suspended particles at the same column, in disagreement with the claimed universal behavior discussed above.

Beyond the surface properties, in practical applications the internal structure of the material (beneath the surface) may be also relevant to determine mechanical and transport properties of the system. In this context, some recent works have focused also on the porous (bulk) structure \cite{Roma,Katz,FDA3,Forg,Kons}. Here, we study the bulk properties of the deposits generated by the BBBD models. Using scaling arguments, we show that the porosity $P$ of the deposits scales as $p^{y-\delta}$ for small $p$, where $\delta$ and $y$ are the exponents defined above (Eq. \ref{EqScaling}). Thus, this work advances over previous ones, which only relate $P$ with $p$ \cite{FDA3,Forg,YuAmar2}, since here we relate directly the scaling exponents of the amplitudes of surface roughness with that of bulk quantities.

The rest of this work is organized as follows. In Section \ref{SecModel} we describe the BD models with bond breaking introduced here. In Section \ref{SecScaling} we present the scaling theory to explain the behaviors of surface and bulk properties for small values of the probability $p$. Numerical results for the BBBDB and BBBDS are presented in Subsections \ref{SecResBBBDB} and \ref{SecResBBBDS}, respectively. Final discussions and conclusions may be found in Section \ref{SecConcl}.

\section{The models}
\label{SecModel}

\begin{figure}[t]
\begin{center}
\includegraphics[width=8.6cm]{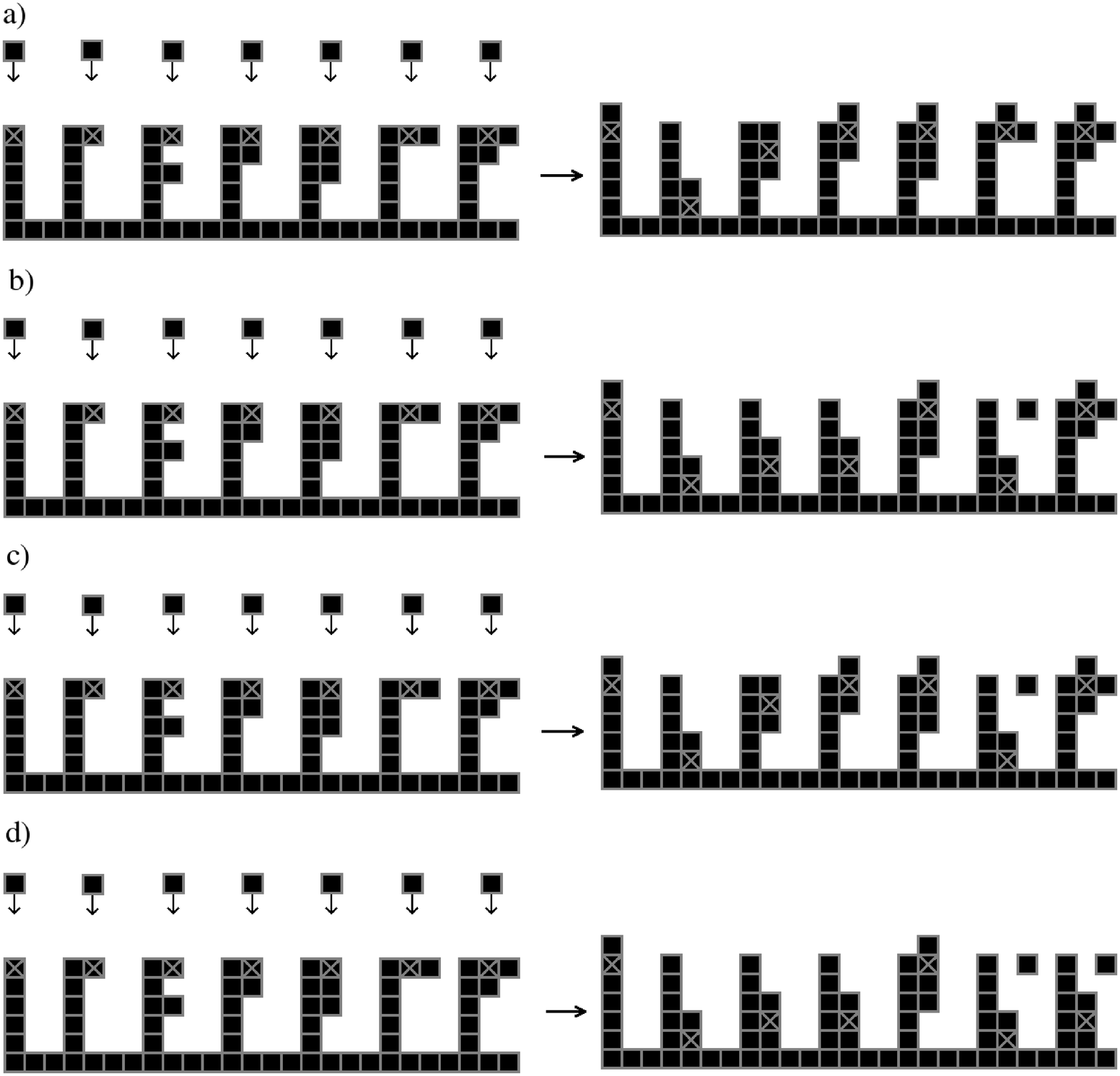}
\end{center}
\caption{Illustration of the bond breaking rules (occurring with probability $1-p$) for the BBBDB models with a) $n=1$ and b) $n=2$, and BBBDS models with c) $n=1$ and d) $n=2$. Deposits before (left) and after (right) deposition of seven particles are shown. The particles which would stay floating due to the broken bonds are deposited following the RD rule before the deposition of the next particle.}
\label{Figgrules}
\end{figure}

In the BBBD models considered here, the impinging particle, following a vertical trajectory, can either dissipate all its energy at collision sticking at the point where it first contacts the surface (BD rule) or transferring energy to the particles below it, dislodging these particles if they are suspended. In this case, the particles (impinging and dislodged) are deposited at the next top below them. This restructuring mechanism could be implemented in two ways: (i) accounting for the number of broken bonds (BBBDB models); and (ii) accounting for the number of dislodged suspended particles  (BBBDS models). In the first case, up to $n$ bonds can break, and, in the second, up to $n$ suspended particles at the same column of the incident particle can be dislodged. In both cases, $n$ gives a measure of the capacity of the incident particle to change the deposit, thus, it is related to the strength of the particle energy. As shown in Fig. \ref{Figgrules}, there are small differences between the two rules, the BBBDS model being more efficient in dislodging particles and therefore in the deposit compaction. 

%When it is not possible to dislodge particles, the impinging particle is deposited at the top of the column of incidence (RD rule). 

In both classes of models, a particle is dropped vertically at a randomly selected column $i$ and sticks on the first site encountered on the surface that is nearest-neighbor of an already deposited particle (BD rule) with probability $p$. In BBBDB models, with probability $1-p$, the number $k$ of bonds of the target (topmost) particle at column $i$ is determined. If $k>n$ the incident particle is deposited at the top of the column, without bond breaking. If $k=n$ the target particle is unbounded and slides down vertically (together with the incident particle) to the next local minimum. If $k<n$, other particles below the target one can be unbounded (and dislodged also), provided that the total number of broken bonds is smaller than or equal to $n$. Figs. \ref{Figgrules} (a)-(b) illustrate this growth rule for $n=1$ and $n=2$.

In BBBDS models, with probability $1-p$, the number $k$ of \textit{consecutive} suspended particles in column $i$ is counted, starting from the top. If $k>n$, the incident particle is deposited at the top of the column, without bond breaking. If $k=n$, the block of $n$ suspended particles and the incident one slide down vertically to the next top. If $k<n$, suspended particles below the first suspended block of particles may also be dislodged, until a block of dislodged particles of size smaller than or equal to $n$ be attained. Figs. \ref{Figgrules} (c)-(d) illustrate this growth rule for $n=1$ and $2$. The particles which would stay floating due to the broken bonds are deposited on the top of the deposit below them, before the deposition of the next particle. However, for small values of $p$, we can just let them be naturally dropped by the (very often) RD particles. We verified that this strategy of simulation, when compared with the procedure of dropping the floating particles after bond breaking, for $p\leq 0.1$, does not affect the values of the crossover times $t_{\times}$, and changes the values of $W_{sat}$ and $P$ in less than $3\%$, which are of the same order as the error bars and, thus, do not change the exponents $\delta$ and $y$.

For $n=0$, the BBBDB and BBBDS models reduce to the RD-BD model proposed by Horowitz and Albano \cite{Horo0}, and it will be referred to here just as the BBBD0 model. Simulations of this model showed that it exhibits a crossover, at a characteristic time $t_{\times}$, from RD to KPZ growth and gave $\delta\simeq 1/2$ and $y\simeq 1$ in $d=1+1$, $d=2+1$, and $d=3+1$ dimensions \cite{Horo0} (see also \cite{Horo1} and references therein). For completeness, we present below simulation results of the BBBD0 model, that are consistent with the results above. Moreover, we present also results for the bulk properties of this model, which were not studied yet. 

\section{Scaling Theory}
\label{SecScaling}

Different scaling arguments have already explained the values of the exponents $\delta$ and $y$ obtained in simulations of several competitive lattice models \cite{Brau,FDA2,Horo1}.
For the BBBD0 model, the basic argument is that, for small values of $p$, the average time for a correlated BD event (which involves lateral aggregation with probability $p$) to take place at a given column is $\tau\sim 1/p$. During the period $\tau$, particles on average are directly deposited onto the surface according to the simple RD rule and the local height increases by $\sqrt{\tau}$.
Thus, the BBBD0 model can be viewed (for small $p$) as a limiting BD model in which
time scales as $\tau\sim 1/p$ while roughness scales as $W\sim \sqrt{\tau}\sim 1/p^{1/2}$.
This explains the scaling function given by Eq. (\ref{EqScaling}) and the conjectured values $\delta=1/2$ and $y=1$ for this model. This argument also gives the relationship $\delta=y/2$ which is valid for all models with crossover from RD to some correlated deposition dynamics \cite{Horo1}.

For other BBBD models, we expect a similar argument. However, due to the bond breaking processes, a single BD event is not sufficient to introduce lateral aggregation and cancel the random fluctuation of the heights. In order to ensure lateral growth, it is necessary that more BD particles fall (consecutively) at the same column. For example, if one suspended particle can be dislodged in the deposit, two consecutive BD events are needed to produce a correlated growth, giving $\tau\sim 1/p^2$ and $W\sim \sqrt{\tau}\sim 1/p$, and the exponents $\delta=1$ and $y=2$. In general, if $n$ bonds can break, $n+1$ ballistic particles are required to introduce correlations, which leads to $\tau\sim 1/p^{n+1}$ and $W\sim \sqrt{\tau}\sim 1/p^{(n+1)/2}$. Thus, we expect the exponents
\begin{equation}
 \delta=(n+1)/2  \quad \quad \text{and} \quad \quad y=(n+1)
\label{EqExponents}
\end{equation}
for these models, in any substrate dimension. We notice that what leads to these values is the number of correlated (BD here) events needed. Thus, if $n$ suspended particles at the same column of the incident particle can be dislodged, again $n+1$ ballistic particles are required to correlate the system and the same exponents are expected. Therefore, the exponents would depend only on the characteristic times, not on the details of the models. It is worth mentioning that in all previously studied competitive models involving RD and solid-on-solid (SOS) correlated models (for example, RD-RSOS (restricted SOS) \cite{FDA2} and RD-RDSR (random deposition with surface relaxation) \cite{Horo1,Horo2}), the exponents $\delta=1$ and $y=2$ were found. In contrast, systems with RD and BD-like models (for example, the BBBD0 model \cite{Horo0,Horo1}) have the exponents $\delta = 1/2$ and $y=1$. This resulted in a classification of the models in two groups: RD-SOS models (with $\delta=1$ and $y=2$) and RD-BD like models (with $\delta=1/2$ and $y=1$) \cite{Kola,FDA2,Kola2,Horo1}, and these two sets of exponents were claimed to be universal (the only two pair of possible values) in random-to-correlated crossover \cite{Horo1}. Our results show that for RD-BD like models, the exponents would be $\delta=(n+1)/2$ and $y=n+1$, which gives $\delta=1/2$ and $y=1$ in the case $n=0$ (no bond breaking) already considered.

It is possible to extend the scaling arguments presented above to explain the behavior of the bulk porosity $P$ of the deposits, for small $p$. 
The porosity $P$ is the fraction of empty lattice sites inside the deposit:
\begin{equation}
P=\frac{V_{P}}{V_{S}+V_{P}}\;\;\; ,
\label{porosity}
\end{equation} 
where $V_{P}$ is the total volume of the pores (vacant sites) and $V_{S}$ is the total volume of the solid (particles) in the deposit.
For small $p$, the pores are narrow in the horizontal direction ($\sim 1$ site width) but vertically high. A typical height difference between neighboring columns is of order $\sqrt{\tau}\sim p^{-\delta}$ for small $p$ and, thus, the lateral aggregation events create pores with volume of order $p^{-\delta}$.
After $V_{S}$ depositions, the average number of pores created is of order $p^{y}V_{S}$, since the characteristic time in which correlations between neighboring columns are built scales as $\tau\sim p^{-y}$. Thus, we obtain $P\sim p^\Delta$, with $\Delta=y-\delta$, which must be valid in any substrate dimension. It is worth mentioning that for the bidisperse BD model (for which $\delta=1/2$ and $y=1$) the scaling behavior $P\sim p^{1/2}$ was obtained in Ref. \cite{FDA3}, consistent with the most general scaling above. Here, we must expect the exponents $\Delta =(n+1)/2$ for the BBBD models.

\section{Numerical Results}

The following results are for simulations on $d=1+1$ substrates of sizes in the range $L=64-1024$, and $d=2+1$, with sizes in the range $L=8-128$. In both cases, we assume periodic boundary conditions.
Averages over $400-1000$ different runs were made.

\begin{figure}[t]
\begin{center}
\includegraphics[width=7cm]{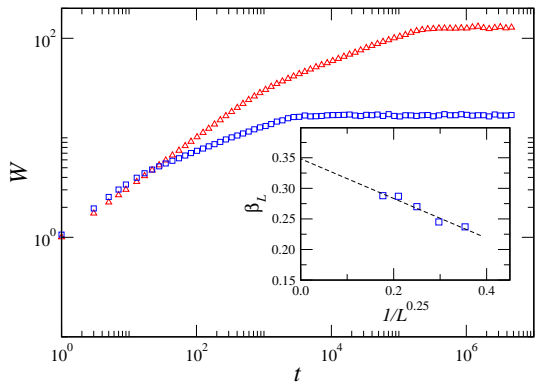}
\end{center}
\caption{(Color online) Roughness $W$ versus time $t$ for the BBBDB1 model with $p=0.05$ (red triangles) and $p=0.5$ (blue squares), for  system size $L=512$. The inset shows effective growth exponents $\beta_{L}$ against $(1/L)^{0.25}$ for $p=0.5$. The linear fit gives the extrapolated value $\beta_{\infty} = 0.35\pm 0.03$.}
\label{FigWversust}
\end{figure}

\begin{figure}[t]
\begin{center}
\includegraphics[width=7cm]{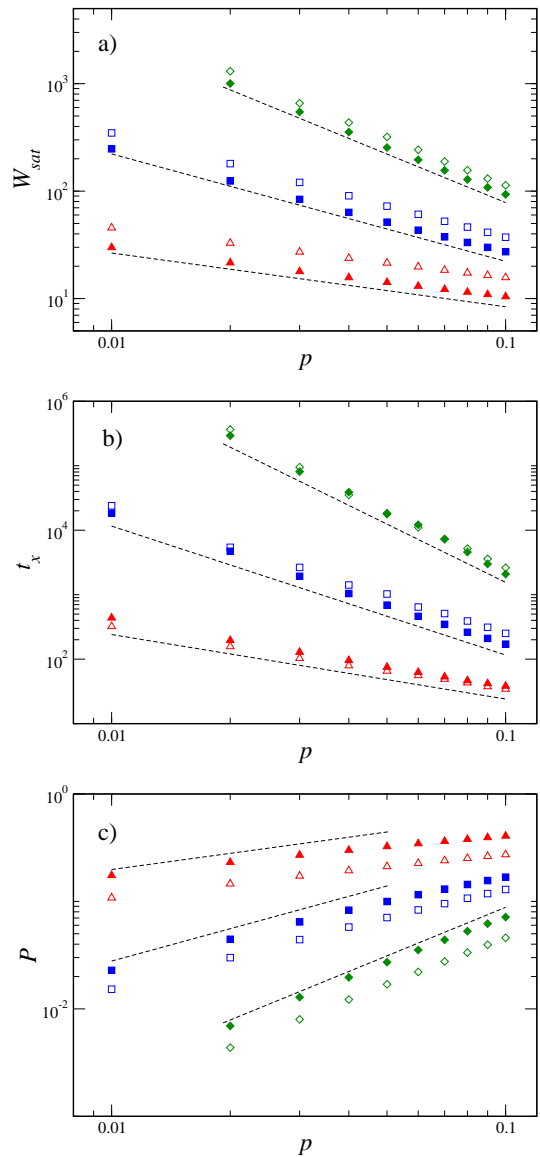}
\end{center}
\caption{a) Saturation roughness $W_{sat}$, b) crossover times $t_{\times}$ and c) porosity $P$ versus probability $p$ for the BBBDB models with $n=0$ (red triangles), $n=1$ (blue squares) and $n=2$ (green diamonds). Data for simulations on $d=1+1$ substrates of size $L=128$ (open symbols) and $d=2+1$ with $L=32$ (full symbols) are shown. In order to improve visualization, in (a) and (c) the data for $n=1$ and 2 in $d=2+1$ are, respectively, multiplied (shifted up) and divided (shifted down) by 1.5. The straight lines have slopes (from the bottom to top): a) $-1/2$, $-1$ and $-3/2$, b) $-1$, $-2$ and $-3$, and c) $3/2$, $1$ and $1/2$.}
\label{FigBBBDB}
\end{figure}

\subsection{BBBDB models}
\label{SecResBBBDB}

Figure \ref{FigWversust} shows plots of the roughness $W$ versus time $t$ obtained for the BBBDB1 model for two values of $p$, in $d=1+1$. For short times, say $t<t_{\times}$, 
the growth is dominated by the RD process and $W\sim t^{1/2}$ for $p<1$. At intermediate times, $t_{\times}<t<t_{S}$, correlations have developed and the BD process dominates giving $W \sim t^{\beta}$. Using data for different lattice sizes and extrapolating to $L\rightarrow\infty$, we obtained growth exponents $\beta \simeq 0.33$ for every studied value of $p>0$ in $d=1+1$. This procedure to obtain $\beta_{L\rightarrow\infty}$ is illustrated in the inset of Fig. \ref{FigWversust}, for $p=0.5$. This asymptotic exponent is consistent with the expected KPZ one ($\beta=1/3$) and with previous simulation results for the BD model in $d=1+1$ \cite{DBS}. Finally, for $t>t_{S}$ the correlations can no longer develop due to the geometrical constraint of the lattice size and saturation occurs. Similar $W \times t$ behaviors were found in all models studied here with $p<<1$ and in other systems with random to KPZ crossover \cite{Horo1,FDA3}.

Below, we present results for the BBBD0 and BBBDB models with $n=1$ and 2, in $d=1+1$ and $d=2+1$ substrates.

In Fig. \ref{FigBBBDB}(a) the saturation roughness $W_{sat}$ as a function of $p$ is shown. As expected, for a given $p$, a larger $n$ implies a larger $W_{sat}$. For small values of $p$, straight lines are observed in the log-log plot, in agreement with Eq. (\ref{EqScaling}), and the best fits give the exponents $\delta$ in excellent agreement with the scaling theory of Sec. \ref{SecScaling} (see Table \ref{TabBBBDB}).

The crossover times $t_{\times}$, for different values of $p$, are shown in Fig. \ref{FigBBBDB}(b). The straight lines in the log-log plot, for small values of $p$, show that the scaling behavior $t_{\times}\sim p^{-y}$ from Eq. \ref{EqScaling} is satisfied and yields exponents $y$ also in good accordance with the predicted ones, as summarized in Table \ref{TabBBBDB}.

As the roughness, the porosity of the aggregate grows in time and attains a stationary value $P$ for large $t$. Fig. \ref{FigBBBDB}(c) shows this stationary porosity as a function of the probability $p$. A good scaling behavior is found for small $p$ and the exponents $\Delta$ obtained are consistent with the expected values (see Table \ref{TabBBBDB}).

The results above are for substrate sizes $L=128$ (in $d=1+1$) and $L=32$ (in $d=2+1$). The system sizes are limited by the saturation times (for $p\ll 1$) which become very long for large $n$. From simulations for other system sizes (smaller and larger in some specific cases) we obtain exponents close to those shown in Table \ref{TabBBBDB}, in accordance with the fact, already pointed out in Ref. \cite{Horo1}, that the exponents $\delta$ and $y$ have weak finite-size corrections. 

Summing up, our numerical results are consistent with the roughness scaling relation (\ref{EqScaling}), and the obtained exponents $\delta$, $y$ and $\Delta$  confirms the exact values from the scaling theory presented here.

\begin{table}[t]
\begin{center}
\begin{tabular}{cccccccc}
\hline\hline
            &  &    BBBD0    &  &    BBBDB1    &  &   BBBDB2   \\
\hline
$y$         &  &  $0.98(3)$   &  &  $1.96(5)$   &  &  $3.0(1)$  & \\
$\delta$    &  &  $0.46(5)$   &  &  $0.98(3)$   &  &  $1.50(2)$  & \\
$\Delta$    &  &  $0.43(3)$   &  &  $0.97(4)$   &  &  $1.48(4)$  & \\
\hline
$y$         &  &  $1.02(5)$   &  &  $2.02(3)$   &  &  $3.03(7)$  & \\
$\delta$    &  &  $0.47(4)$   &  &  $0.98(3)$   &  &  $1.49(3)$  & \\
$\Delta$    &  &  $0.40(5)$   &  &  $0.96(5)$   &  &  $1.50(3)$  & \\
\hline\hline
\end{tabular}
\caption{Amplitude exponents $y$ and $\delta$ and porosity exponent $\Delta = y-\delta$ for $d=1+1$ (top) and $d=2+1$ (bottom).}
\label{TabBBBDB}
\end{center}
\end{table}

\subsection{BBBDS models}
\label{SecResBBBDS}

We made simulations of BBBDS models with $n=1$, 2 and 3 in $d=1+1$ and $d=2+1$ dimensions. As discussed above, the evolution of the roughness  with time has a behavior similar  to the one shown in Fig. \ref{FigWversust}, for all BBBDS models with $p<<1$. 

\begin{figure}[t]
\begin{center}
\includegraphics[width=7cm]{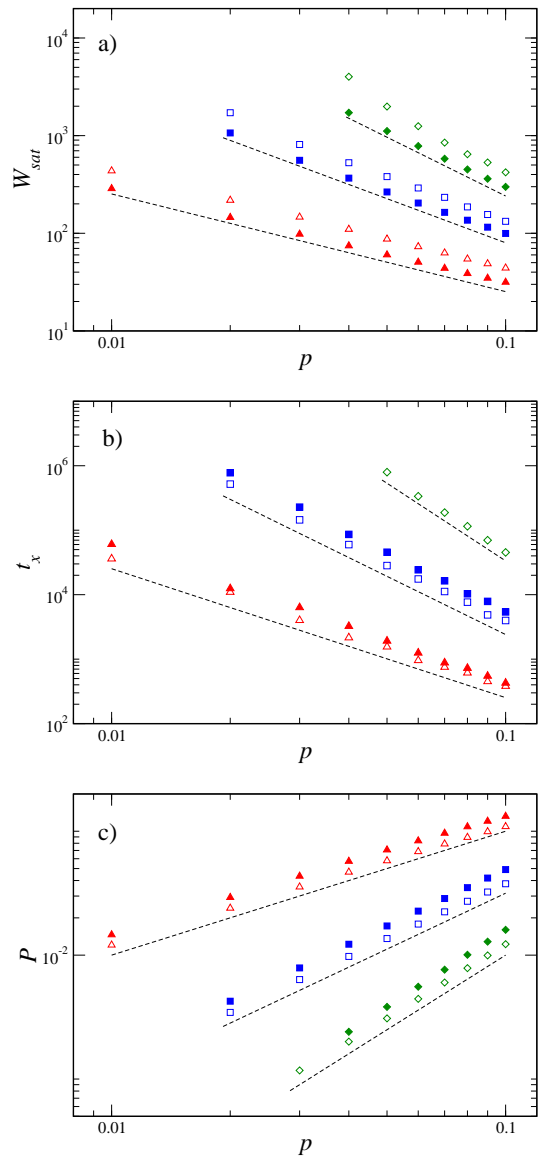}
\end{center}
\caption{a) Saturation roughness $W_{sat}$, b) crossover times $t_{\times}$ and c) porosity $P$ versus probability $p$ for the BBBDS models with $n=1$ (red triangles), $n=2$ (blue squares) and $n=3$ (green diamonds). Data for simulations on $d=1+1$  substrates (open symbols) of size $L=128$ and $d=2+1$ (full symbols) with $L=32$ (for $n=1$ and 2) and $L=16$ ($n=3$) are shown. In order to improve visualization, in (c) the data for $d=2+1$ are divided (shifted down) by 1.4. The straight lines have slopes (from the bottom to top): a) $-1$, $-3/2$ and $-2$, b) $-2$, $-3$ and $-4$, and c) $2$, $3/2$ and $1$.}
\label{FigBBBDS}
\end{figure}

In the BBBDS models with a given $n$, more particles are dislodged than in the BBBDB ones (see Fig. \ref{Figgrules}). Thus, in the former the correlations develop more slowly,  making the crossover times and the saturation roughness larger. On the other hand, the porosity is smaller in BBBDS models, since more particles are compacted by dislodgement. For example, for $n=1$ and fixed $L$ and $p$, we found $W_{sat}^{BBBDS}/W_{sat}^{BBBDB} \simeq t_{x}^{BBBDS}/t_{x}^{BBBDB} \simeq 1.5$ in $d=1+1$ and $\simeq 3$ in $d=2+1$, while $P^{BBBDB}/P^{BBBDS} \simeq 1.2$ in $d=1+1$ and $\simeq 1.5$ in $d=2+1$.

In Fig. \ref{FigBBBDS} we show the saturation roughness $W_{sat}$, the crossover time $t_x$ and the stationary porosity $P$ as functions of $p$. As in BBBDB models, we found here good scaling behaviors, for small $p$, consistent with the scaling relation given in Eq. \ref{EqScaling}. The exponents $y$, $\delta$ and $\Delta$ are in good agreement with the expected ones, as shown in Table \ref{TabBBBDS}. These results confirms the correctness of our scaling theory and shows that the exponents are in fact universal (independent on model details), as discussed in Sec. \ref{SecScaling}.

The data above are for substrate sizes $L=128$ in $d=1+1$. In $d=2+1$ we set $L=32$ for $n=1$ and 2, and $L=16$ for $n=3$. Exponents obtained for other system sizes are close to the ones in Tab. \ref{TabBBBDS},  showing that such exponents have negligible finite-size effects. For $n=3$, the large crossover and saturation times limit our simulations to substrate sizes $L\leq 16$ in $d=2+1$. For these sizes, there is not a clear KPZ region in the $W \times t$ curves: the crossover and saturation times became very close, so that it is not possible to determine the former one. Thus, for this case, we  obtained the exponent $y\simeq 3.95$ from the relation $\Delta = y - \delta$.

\begin{table}[t]
\begin{center}
\begin{tabular}{cccccccc}
\hline\hline
            &  &    BBBDS1    &  &    BBBDS2    &  &   BBBDS3   \\
\hline
$y$         &  &  $2.01(4)$   &  &  $3.04(5)$   &  &  $4.0(1)$  & \\
$\delta$    &  &  $0.99(1)$   &  &  $1.50(2)$   &  &  $2.0(1)$   & \\
$\Delta$    &  &  $0.98(4)$   &  &  $1.49(2)$   &  &  $1.98(3)$  & \\
\hline
$y$         &  &  $2.1(1)$    &  &  $3.05(5)$   &  &  $--$  & \\
$\delta$    &  &  $0.97(4)$   &  &  $1.50(2)$   &  &  $1.93(5)$  & \\
$\Delta$    &  &  $0.99(3)$   &  &  $1.52(2)$   &  &  $2.02(4)$  & \\
\hline\hline
\end{tabular}
\caption{Amplitude exponents $y$ and $\delta$ and porosity exponent $\Delta = y-\delta$ for $d=1+1$ (top) and $d=2+1$ (bottom).}
\label{TabBBBDS}
\end{center}
\end{table}

\section{Discussions and Conclusions}
\label{SecConcl}

We studied the effects of the surface restructuring mechanism on roughness scaling and bulk properties in ballistic-like models in which the bond breaking of suspended particles is allowed (the BBBD models). We found that the dislodgement of previously aggregate particles do not change the asymptotic universality class of the system, i. e., for $p>0$ the scaling exponents $\alpha$, $\beta$ and $z$ are in the expected KPZ class. However, the restructuring mechanism produces a crossover from random to KPZ growth in the roughness evolution that follows the same scaling relation observed in previous models with random to correlated growth. Using scaling arguments we show that the exponents describing the roughness and crossover time amplitudes, for the BBBD models, are $\delta = (n+1)/2$ and $y=(n+1)$, where $n$ could be either the maximal number of broken bonds or of dislodged suspended particles at the same column of the incident particle. Such exponents are confirmed by numerical simulations on one- and two-dimensional substrates (summarized in Tables \ref{TabBBBDB} and \ref{TabBBBDS}) and they are expected to hold in any dimension. These results generalize the classification present in literature, in which models with random-to-correlated competition were divided into two groups: solid-on-solid ones, with $\delta = 1$ and $y=2$, and ballistic-like ones, with $\delta = 1/2$ and $y=1$. These exponents are not universal, as claimed in previous works, and in ballistic-like models we have the general exponents found here (Eq. \ref{EqExponents}).

Models accounting for the maximal number of broken bonds (BBBDB) and for the maximal number of dislodged suspended particles (BBBDS) have different rates of particle dislodgement, which leads to large differences in crossover times, saturation roughness and porosity. However, they have the same exponents (given by Eq. \ref{EqExponents}), showing that these ones are independent of details of the model, depending only on the characteristic times involved, as expected from our scaling theory. Thus, we can conjecture that in every system where $n+1$ correlated events are needed to introduce correlations, the exponents of the BBBD models (Eq. \ref{EqExponents}) will be found.

Finally, as expected, the restructuring mechanism leads to deposits with more compact bulks. Very interestingly, the bulk porosity $P$ is related to the probability $p$ (for small $p$) through the surface exponents discussed above, with $P \sim p^{y-\delta}$. This scaling relation was confirmed with numerical simulations on $d=1+1$ and $d=2+1$ substrates, and must be valid in any substrate dimension. It also explains the relation $P \sim p^{1/2}$ found for the bidisperse BD model in \cite{FDA3}, and must be valid for any system with random-to-correlated crossover forming a porous deposit.

\acknowledgments

We would like to thank Sukarno O. Ferreira for helpful discussions and the support from Capes, CNPq and FAPEMIG (Brazilian agencies).

%----------------------------------------------------------------------------%
% -------------------------- LISTA DE REFERENCIAS ---------------------------%
%\newpage

% ------------------- Fim das Referencias Bibliograficas --------------------%

      \end{document}